\documentclass{WileyMSP-template}
\usepackage{graphicx}      
\usepackage{amsmath}       
\usepackage{setspace}      
\usepackage{fancyhdr}      
\usepackage{geometry}      
\usepackage{tocbibind}     
\usepackage{appendix}      
\usepackage{cite}          
\usepackage{hyperref}      
\usepackage{titlesec}      
\usepackage{multirow}      
\usepackage{graphicx}      
\usepackage{array}
\usepackage{hyperref}
\usepackage{float}
\usepackage{listings}
\usepackage[hyphenbreaks]{breakurl}
\usepackage{algorithm}
\usepackage{algpseudocode}
\usepackage[table,xcdraw]{xcolor}
\usepackage{float}
\begin{document}

This work has been submitted to Wiley for possible publication. Copyright may be transferred without notice, after which this version may no longer be accessible.

\newpage

\pagestyle{fancy}

\title{Enhancing Tennis Training with Real-Time Swing Data Visualisation in Immersive Virtual Reality}

\maketitle

\author{Ryan Najami}
\author{Rami Ghannam*}

\begin{affiliations}
Mr Ryan Najami and Prof. R. Ghannam\\
James Watt School of Engineering, The University of Glasgow\\
Email Address: 2545106N@student.gla.ac.uk\\
Email Address: rami.ghannam@glasgow.ac.uk
\end{affiliations}



\begin{abstract}

Recent advances in immersive technology have opened up new ways in sports training, particularly for activities requiring precise motor skills such as tennis. In this paper, we demonstrate a virtual reality (VR) tennis training system that integrates real-time performance feedback via a wearable sensor device. Ten participants were invited to wear the sensing device on their dominant hand to capture motion data such as swing speed and swing power while engaging in a VR tennis environment. During a baseline test, participants played without access to these metrics. In a subsequent test, the same participants performed similar routines but with their swing data displayed in real time on a floating VR overlay. The qualitative and quantitative findings revealed that including real-time visual feedback contributed to improved shifts in performance behaviors and enhanced players’ situational awareness. Some participants demonstrated improved swing consistency and strategic decision making, although changes in precision varied between individuals. Furthermore, subjective feedback indicated that the immersive experience, combined with instantaneous statistical insights improved engagement and motivation. These findings demonstrate the value of VR-based data visualisation as a training tool, suggesting broader applications in sports performance.


\end{abstract}


\section{Introduction}
Recent advances in technology have significantly changed how athletes train, especially in sports requiring precise movements like tennis. Virtual reality (VR) now allows players to practice without needing physical tennis courts or traditional equipment. Research shows VR sports environments are engaging and enjoyable, motivating players to exercise more regularly \cite{norGamification2019}. VR tennis training has become particularly popular because it offers realistic and interactive experiences, improving player motivation and training effectiveness \cite{AndelVRtennis}.
\vspace{3mm}

Sensor-based tennis training has existed since Hawk-Eye was introduced in 2002 \cite{hawkeyecite}. Following this, new technologies such as Babolat Play have emerged, embedding sensors directly into tennis rackets to track swing data \cite{BabolatPlay}. More recent solutions include sensors integrated in tennis balls \cite{GhannamTennisBall2024}, or mounted on racket handles such as the STA 4.0 Smart Tennis Analyzer \cite{STAswinganalyser}, or attached to racket strings, as seen with the Qlipp sensor dampener \cite{qlippsensorcite}. Wearable devices like Babolat POP have also been used to monitor player movements \cite{BabolatPopcite} (discontinued). These devices typically use inertial sensors to measure racket and arm movements.
\vspace{3mm}

However, current VR tennis training systems mainly simulate gameplay and often lack detailed real-time feedback on critical metrics such as swing speed and power. Novak et al. demonstrated VR training could improve players' dynamic balance, but there is limited research examining whether displaying detailed performance data directly improves tennis skills \cite{norGamification2019}.

\vspace{3mm}

The goal of this study is to demonstrate if beginner tennis players can improve their skills by viewing their performance data in real-time during VR training. By integrating sensors that track movement and clearly displaying this information inside the VR environment, we aim to verify if instant feedback helps players achieve more consistent swings and better decision-making during play. In fact, our hypothesis is that real-time visualisation of swing metrics (speed and power) during VR tennis training will lead to a statistically significant increase in the number of points scored against an AI opponent for novice players compared to a baseline condition without visualization

\vspace{3mm}

The remainder of this paper is structured as follows: Section 2 provides a review of the literature covering immersive technology applications, data visualization techniques, sensor positioning, and current research gaps. Section 3 describes our methodology, including hardware specifications, software integration, and experimental setup. Section 4 discusses the implementation of the training and testing procedures, while Section 5 presents our results. Finally, Sections 6 and 7 discuss findings and offer conclusions, highlighting the implications for future research and practical applications.


\section{Literature Review}

\subsection{Applications of Immersive Technology in Sports}
Virtual Reality (VR) technology is increasingly being used as a tool to complement training practices for users of all skill levels. In recent years, it has become common for athletes and novice sports players to simulate the experience of playing a real sport by immersion in a virtual environment. Performance data metrics can be taken from the simulation and evaluated to provide the user with a quantifiable understanding of training progression. Highlighted by Kumar (2020), the use of extended virtual reality practices in the context of physical training was shown to increase athletic training in students, while simultaneously benefitting their learning ability \cite{kumar2022gamification}. From immersive technology combined with data collection, players receive a unique immersive experience with a perspective to analyse their performance, and therefore make necessary adjustments to improve. While the implementation of immersive technology in tennis training is still in its early stages, and is considered relatively recent in the context of skill development, it can be an effective tool in helping players practice the fundamental skills required to develop a high level of proficiency \cite{NeumannVRreviewtennis}. 

\subsection{Data Visualisation}
Current research shows that direct visualisation of key virtual parameters can positively impact the performance of the player. Wu et al. (2021), examines how various aspects of user experience when immersed in a gamified virtual environment are based on several factors, that when well defined, trialled, and monitored, can improve their performance considerably \cite{Wu2021SPinPong}. Embedding a training environment with engaging performance visuals provides the player with valuable information, which can create incentives to adapt training techniques to improve results.
Cossich et al. (2023), discusses the transformative impact data visualisation has on current sports performance analysis techniques \cite{CossichTechBreakthroughs}. By simplifying complex data, visual representation can provide more insightful information to players, resulting in better decision-making. For tennis, insightful data can be attributed to information regarding tennis swings. Polk et al.(2020) presented an approach towards visualised tennis data in which serve speed and stroke type were included as collected metrics \cite{PolkTennisVisAnalytics}.
visualised data metrics can enable researchers to monitor behaviour patterns quantitatively. Wallner et al. (2013) mentions that the use of numerical gameplay data has become a valuable asset for performance analysis \cite{wallner2013visualization}. The article continues to highlight the various iterations of data visualisation instruments: charts and graphs, numerical displays and heat maps were identified.

\subsection{Data Acquisition and Sensor Positioning}
As demonstrated by Kos et al. (2016), the use of wearable IMU devices is advantageous for acquiring valuable data related to tennis strokes \cite{KosTennisStroke}. The IMU used is a MEMS (Micro-Electro-Mechanical System) and includes a 3-axis gyroscope and 3-axis accelerometer, providing 6 degrees of freedom (DOF). Highlighted in the paper, the 6-DOF components retrieve valuable data which, in the context of tennis strokes, has been used to detect and classify whether a forehand, backhand, or a serve has been played. The device comprises the IMU, microcontroller, and power supply (Lithium-Ion Polymer Battery). In the context of improving performance by metric visualisation in real-time, the data obtained from such a device can be interpreted and adapted to provide insightful information for the player.

\vspace{3mm}

Obtaining accurate swing data considers the location of the IMU sensor. To best retrieve accurate data, the sensor positioning must emulate that of the arm or racket orientation. A study from Ebner et al. (2020), compared the accuracy of stroke detection and classification for 8 different stroke types in the case of two sensor locations: 1. Located on the racket and 2. Located on the wrist \cite{EbnerIMUClassification}. The results of the study showed similar performance for accuracy in stroke detection and classification for both cases, indicating that, provided the sensor remains in line with the range of swing motion, data will not be biased towards either of the mentioned sensor locations. This relaxes the conditions required to fabricate a device used to collect swing data. The implementation of a wireless sensor device can be interpreted for various iterations, without compromising the accuracy of results.

\subsection{Limitations and Research Gap}
Limitations of immersive technology in the context of training have been researched. Highlighted by Le Noury et al. (2022), the application of VR for training lacks the realism associated with engaging in the real sport: with the absence of ball contact and dynamic movement on the court, this training method falls short of providing a complete experience, which can be overcome by direct visualisation to result in continuous development and optimised learning \cite{LeNouryXRstate}. Velev et al. (2017) highlights the weakness that VR is often considered as a game and is not taken seriously in the context of education and teaching, which can be applied to sports training \cite{velev2017virtualchallenges}. This research indicates that some users cannot adapt to virtual training scenarios, despite promising applications of VR technology in several related fields.

\vspace{3mm}

Nor et al. (2019) revealed that research findings lack variety in wider demographics\cite{norGamification2019}. The literature on the performative applications of VR sports training is limited \cite{MichalskiVRtabletennis}, and so this leaves a noticeable gap in research on the impact of data visualisation on performance in a virtual tennis training environment for novice players.  Existing tennis VR systems may focus on game simulations, but lack comprehensive integration of detailed statistical data such as swing speed, force vectors, or detailed shot patterns. In addition, there is limited research on how to incorporate visualisation that is comprehensible and actionable for a VR tennis game and how to define effective visualisation metrics to improve decision-making or performance for players \cite{LeNouryVRTennis}.

\section{Methodology}

This section discusses the necessary requirements to complete the comprehensive study on the impact of data visualisation on novice tennis players. Our study incorporates a mix of hardware and software elements in parallel with participant interaction. In the following section we outline hardware design, device procurement and use and software logic. 

\subsection{Specification}
There are many existing technologies used in tennis to track swing data, each with their own unique features. The swing data shown to players must be comprehensive enough to provide sufficient insight into player performance. Two metrics were chosen for collection and visualization from a gyroscope: swing speed and swing power. Swing speed is widely tracked by existing tennis sensor technologies, offering users direct performance feedback. Swing power, also a common metric in commercial tennis products, was selected as the second metric because it effectively complements the speed data.
\vspace{3mm}

Swing speed was calculated by first obtaining angular velocity in degrees per second (deg/s). A gyroscope typically tracks the sensor's rotational movement along each axis: Roll (X), Pitch (Y), and Yaw (Z), providing angular velocity for each axis. 

\begin{equation}
    \omega \, (\text{deg/s}) = \sqrt{\text{gyro.x}^2 + \text{gyro.y}^2 + \text{gyro.z}^2}
\end{equation}

This value converts into radians per second (rad/s) by multiplying by $\frac{\pi}{180}$:
\begin{equation}
    \omega(\frac{rad}{s})=\omega(\frac{deg}{s})*\frac{\pi}{180}
\end{equation}

Linear velocity, in meters per second, was found from the following formula:
$v=r\omega$, where $r$ = radius, and $\omega$ = angular velocity. Here, the radius of the swing was taken as the length of the racket, assumed to be 0.68 m. Swing speed was obtained in miles per hour by multiplying linear velocity by a conversion constant of 2.23694 (m/s to mph).

\vspace{3mm}

Swing power was shown as a normalised value of the maximum kinetic energy of the swing. The calculation combines linear and rotational component of kinetic energy to achieve a total raw power value: the "power" of the current swing. As this is a normalised value of kinetic energy, the term "power" was not used as a conventional meaning with units, rather as a term used generally to describe how forceful the swing was compared to the players' maximum value, as do current technologies.
\vspace{3mm}

Linear and rotational kinetic energy were found using the following formulae:

\begin{equation}
    E_k{\text{(linear)}} = \frac{1}{2} m v^2 \;\;and \;\;E_k {\text{(rotational)}} = \frac{1}{2} I \omega^2
\end{equation}
\vspace{3mm}

Where $m$ = Effective mass of racket (kg), $v$ = linear velocity ($\frac{meters}{second}$), $I$ = Racket moment of inertia ($\frac{kg}{m^2}$) and $\omega$ = Angular velocity ($\frac{rad}{s}$)

The effective racket mass was assumed to be $0.2kg$ and the moment of inertia was assumed to be $0.045kgm^2$.
\vspace{3mm}

Total "raw power" was the sum of the two components:

\begin{equation}
    P_{\text{raw}} = E_k{\text{(total)}} = E_k{\text{(linear)}} + E_k{\text{(rotational)}}
\end{equation}
\vspace{3mm}

Normalisation begins with a tracking of the minium and maximum power values which are stored in variables {\fontfamily{pcr}\selectfont minPowerObserved} and {\fontfamily{pcr}\selectfont maxPowerObserved}.
\vspace{3mm}

Raw power is normalised after {\fontfamily{pcr}\selectfont minPowerObserved} and {\fontfamily{pcr}\selectfont maxPowerObserved} are updated:
\begin{equation}
    P_{\text{normalised}} =\frac{P_{\text{raw}} - P_{\text{min}}}{P_{\text{max}} - P_{\text{min}}} * 100
\end{equation}
\vspace{3mm}

The final value was constrained between 0-100 to provide a percentage of the output.
Maximum swing power is stored from the first swing, which is recorded as 100\%. Subsequent swings are then normalised based on this initial swing. In the event a subsequent swing exceeds the initial swing, the max swing power value updates and every subsequent swing after that is based on the new power value.
\vspace{3mm}

This calculation is based on a derivation of the work done throughout a swing from Racket Research \cite{RacketResearch}. The calculation incorporates the use of effective mass which is based on a calculation from Impacting Tennis \cite{ImpactingTennis}. As the calculation incorporates data collected from the sensor, the output value was true to each swing.
\vspace{3mm}

The swing speed calculation had to be validated by experimental measures in order to confirm that data shown to players was correct. Validation was carried out through an experimental process, in which the device was manually moved some distance and timed. The speed was found from a distance/time calculation. The result of this calculation was compared with the result from the sensor device. 
\vspace{3mm}

The data collected from the device showed a 1:1.15 accuracy ratio compared to the result obtained from manual testing. In the code, a calibration factor of 1.15 was applied to the swing speed calculation, keeping it in line with the manual test data.
\vspace{3mm}

Initially, we attempted to mount all hardware components directly onto the existing VR controller. However, the controller's limited surface area proved insufficient to house the components. The final design incorporates a wearable sleeve with integrated pouches for the hardware. This approach allows players to hold the VR controller while maintaining adequate wrist flexion.

\subsection{System Design and Components}


Facilitating the visualisation of technical data required a combination of hardware and software components. To obtain realistic racket movement data, the sensor used should allow sufficient raw data to be taken and interpreted into appropriate metrics. The final value for swing speed was displayed in miles per hour (mph) to align with common speed data used in tennis. Communication between the sensor and computer was facilitated using a microcontroller.
\vspace{3mm}

The equipment used should be unobstructive to player movement, with components covering a small surface area and having minimal weight. To remove the restriction of wired connections, the device was designed with a portable battery power supply and sensor data transmitted wirelessly to a computer. Additionally, the cost of hardware was considered. It was important to choose low-cost items to facilitate the cost-effective replacement of potential damaged items.

\subsection{Hardware Components}
\begin{center}
    \begin{figure}[h!]
    \centering
        \fbox{\includegraphics[scale=0.3]{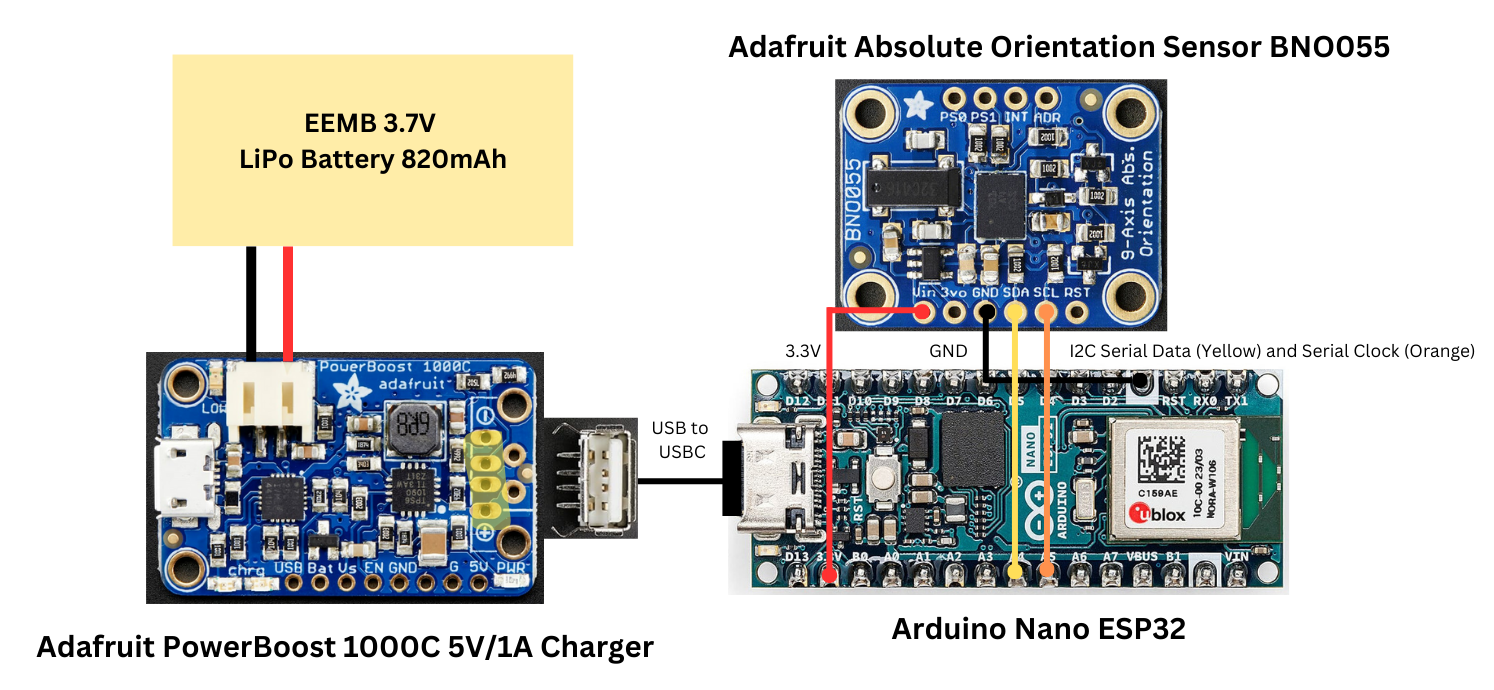}}
        \caption{Component connection diagram - 3.7 V LiPo battery connects to the PowerBoost Charger using a JST connector. The PowerBoost Charger provides a stable 5 V to the Arduino Nano ESP32 using a USB-A to USB-C cable. Highlighted in yellow is the area for the soldered USB-A port. I2C communication occurs between pins A4 and SDA (Yellow) and A5 and SCL (Orange). 3.3 V pin supplies 3.3 V to the BNO055 module (red), and ground connection is made between GND pins (black).}
        \label{fig:sensor connection}
    \end{figure}
\end{center}

The set-up includes a VR headset, controllers, and a VR tennis game. For this study, a Meta Quest Pro VR headset and controllers were used in conjunction the VR game First Person Tennis by Mikori Games \cite{FirstPersonTennis}. While the choice of headset is not strictly limited to the Meta Quest Pro, the VR tennis game had to be selected carefully. The final implementation of the design for data visualisation required an interface with a desktop PC, which required a tennis game that is PC-VR compatible. The device incorporates an IMU sensor, microcontroller, and power supply, the details of which are outlined below. The specific hardware components chosen to develop the prototype sensor device are detailed in this section. Figure \ref{fig:sensor connection} provides a visual diagram illustrating the connections between each component.

\subsubsection{Microcontroller}
The Nano ESP32 from Arduino \cite{NanoESP32cite} was selected as it is a small, compact microcontroller featuring wireless connectivity via Wi-Fi and Bluetooth. It measures 43.1 x 17.78 mm and weighs 30 grams, and is priced at £19.90, making it a suitable option. The wireless connectivity options allow for a portable design, where sensor data can be sent wirelessly to a computer.

\subsubsection{Sensor}
The Bosch BNO055 Intelligent 9-axis Smart Sensor \cite{BoschBNO055cite} was chosen as the IMU for the device. The sensor contains a 3-axis accelerometer, 3-axis gyroscope and 3-axis magnetometer, making it 9-DOF. The Adafruit BNO055 Absolute Orientation Sensor Module \cite{AdafruitBNO055modulecite}houses the smart sensor and obtains raw data from an accelerometer, gyroscope and magnetometer, which can be interpreted to obtain insightful swing data. The sensor is priced at £34.43 and weighs 3 grams. Moreover, the sensor can implement sensor fusion technology, which combines data from all three sensors to achieve a more optimised result.

\subsubsection{Power Supply}
The EEMB Lithium Polymer battery 3.7 V 820 mAh 603449 LiPo Rechargeable Battery Pack \cite{EEMBLiPocite} and Adafruit PowerBoost 1000 Charger - Rechargeable 5 V LiPo USB Boost @ 1 A - 1000 C \cite{PowerBoost} were used. The Nano ESP32 required a steady 5 V power supply for operation. The LiPo battery provides a 3.7 V supply and was used in conjunction with the PowerBoost Charger to output 5 V. The module also features a battery low indicator \cite{ESP32Usermanual}. 
A USB-A port soldered onto the PowerBoost Charger enabled a USB-A to USB-C connection for powering Nano ESP32 with 5 V. The EEMB LiPo battery (£5.39) and PowerBoost Charger (£23.79) weigh a combined 28 grams.

To receive sensor data from BNO055, module uses I2C communication. On the Nano ESP32, pins A4 (serial data - yellow) and A5 (serial clock - orange) were connected using soldered electrical wires to the BNO055 module's SDA and SCL pins. SDA is the data line that transfers data between the BNO055 and Nano ESP32 and SCL is the clock line that synchronises the timing of data transfer. The Nano ESP32 provides 3.3V to the BNO055 module through the 3.3V pin (red). A ground connection (black) was made between the GND pins on both boards.
For stable operating conditions, the documentation for the Nano ESP32 from Arduino recommends a steady supply of 5V. A 15 cm USB-A-to-USB-C cable (£5.99) was connected from a soldered USB-A port on the PowerBoost Charger to the USB-C port on the Nano ESP32. This cable ensured the design could be modified so that adjustments to the system could be made if required. If a permanent connection between the boost charger and Nano ESP32 was made, e.g. soldered wires, this may have raised issues when mounting the kit. The 3.7 V LiPo battery was connected to the PowerBoost Charger using a JST connector, which enabled the battery to be easily removed. The LiPo battery can be recharged from a Micro-USB connection to the PowerBoost Charger while connected.

\subsection{Software}
The software used to facilitate data collection, interpretation, and visualisation using the Arduino Nano ESP32 is outlined. Program development and data collection was facilitated through Arduino applications. The desktop Arduino integrated development environment (IDE) was used for program development, in which the appropriate libraries are readily made available for installation. The Arduino Cloud \cite{Cloudcite} allows for wireless data collection and visualisation. As an Arduino device, setting up the Nano ESP32 with Arduino applications was simplified, resulting in a smoother setup process. The Arduino Nano ESP32 was programmed using C++ language, which is compliant with Arduino hardware.
\vspace{3mm}

\begin{figure}[h!]
        \centering
        \fbox{\includegraphics[scale=0.48]{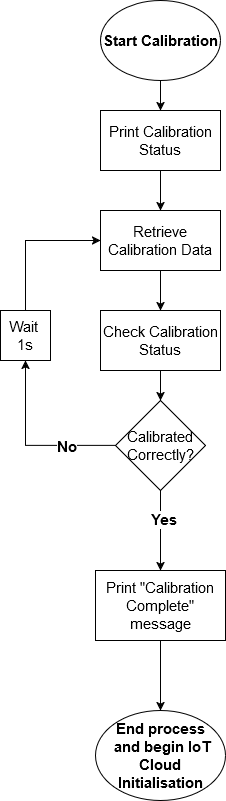}
        \hspace{1in}
        \includegraphics[scale=0.48]{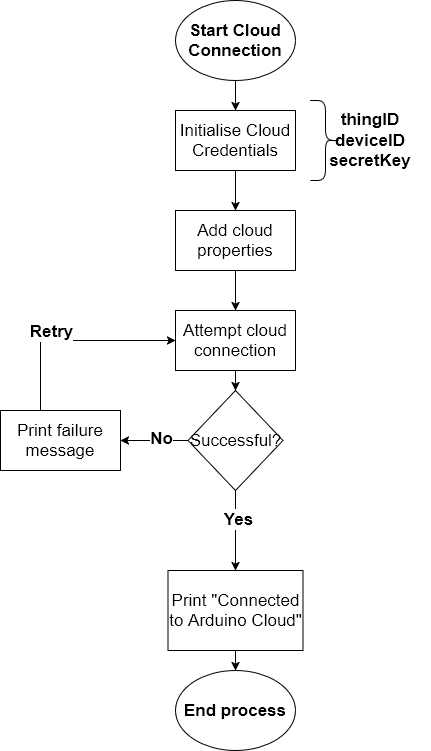}}
        \caption{BNO055 Calibration (Left) and Arduino Cloud Connection (Right) Flowcharts. Calibration is achieved by moving the BNO055 module in specific axis patterns. The program checks for the calibration status before initialising Arduino Cloud connection. Cloud credentials are initialised before attempting a connection. The process finishes once the Cloud connection is achieved.}
        \label{fig: Calibration and IoT flow}
\end{figure}

The Adafruit BNO055 sensor module was calibrated before conducting the Cloud connection. The BNO055 calibration and Cloud connection flowcharts are shown in Figure \ref{fig: Calibration and IoT flow}. 
Stated by documentation from Adafruit's Website: ``The four calibration registers - an overall system calibration status, as well individual gyroscope, magnetometer and accelerometer values - will return a value between `0' (uncalibrated data) and `3' (fully calibrated). The higher the number the better the data will be''. The website explains that the gyroscope is fully calibrated by ``leaving the module still in any position'', ``a figure 8 pattern movement is required'' to fully calibrate the magnetometer, and the accelerometer was fully calibrated by ``placing the module in 6 standing positions for +X, -X, +Y, -Y, +Z and -Z'' \cite{AdafruitDocumentation}.

\vspace{3mm}

\begin{figure}[h!]
        \centering
        \fbox{\includegraphics[scale=0.4]{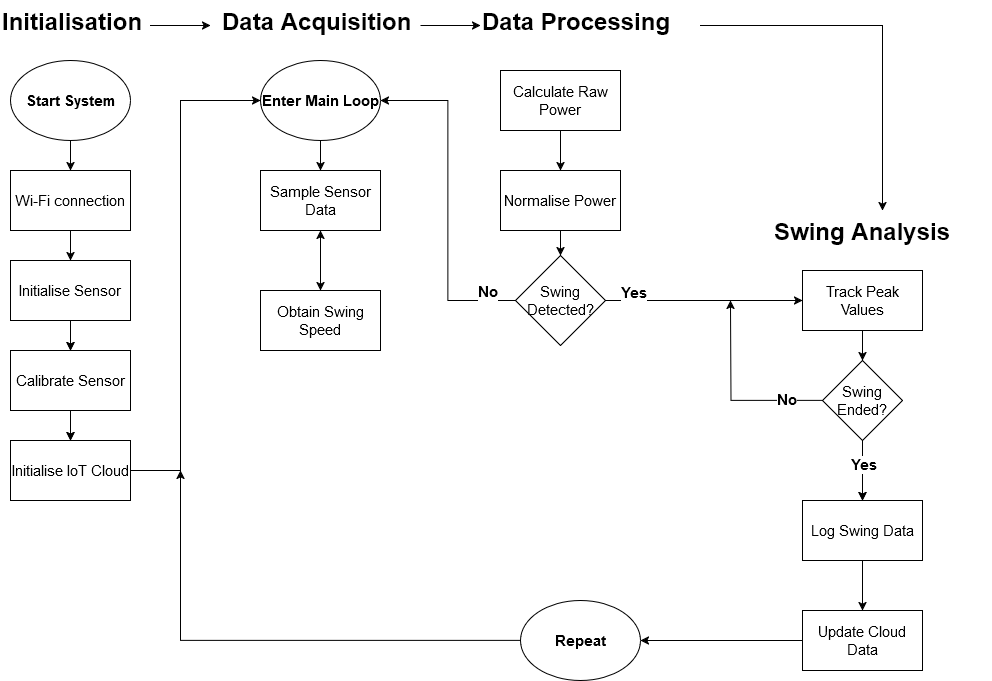}}
        \caption{Complete software flowchart - Data acquisition begins after Wi-Fi connection, calibration, and cloud connection are achieved. The program enters the main loop if a valid swing has been detected. Sensor data is sampled to obtain swing speed and then interpreted to calculate normalised swing power. After each swing, peak values are logged and the Cloud is updated. The process repeats until ended by the user.}
        \label{fig:flowchart}
\end{figure}

The final logic flowchart for the entire system can be seen in Figure \ref{fig:flowchart}. The IDE outputs data in the serial monitor only if the device is connected to the computer via a serial port. Due to the portable demand of the device, data must be collected wirelessly. 
\vspace{3mm}

The code was structured to allow the device to connect to the Arduino IoT Cloud, which is a platform that uses Wi-Fi to send and receive data wirelessly. The Nano ESP32 was setup as a device on the Cloud platform with its own Device ID. Wi-Fi credentials were configured with the device and were stated in the microcontroller code. The device has an associated ``Thing'', in which Variables such as swing speed and swing power can be defined.

\subsection{Participant Selection}
The experience of the participant was an important aspect of the testing process to consider, and therefore select procedures were carried out to ensure players could perform without external adverse influences. 
Participant selection was carried out with the intention of securing novice level players with no experience playing real tennis. The potential for the greatest improvement in performance occurs in the early stages of implementation. When players are just beginning to experience the process of playing an unfamiliar sport, exposure to a new training environment will offer greater disparity in performance results; hence, novice players were desired. In contrast, if intermediate or advanced players were chosen for this study, testing can result in a negligible improvement in performance.
\vspace{3mm}

The selected participants were right-handed 18-24 year old healthy males with little to no experience in tennis. All ethical implications were considered and an ethical approval application was submitted.

\section{Implementation}

This section contains the experimentation phase of the project. Outlined are the selection of participants, the testing framework, sensor data integration and visualisation, and the participant questionnaire.

\subsection{Testing Framework}

To effectively evaluate the effect of data visualisation, an initial baseline test was compared with a visualisation test. Both tests were carried out in an area of appropriate room temperature, and with enough space for a comfortable playing experience. 
\vspace{3mm}

\begin{figure}[H]
    \centering
    \includegraphics[scale = 0.75]{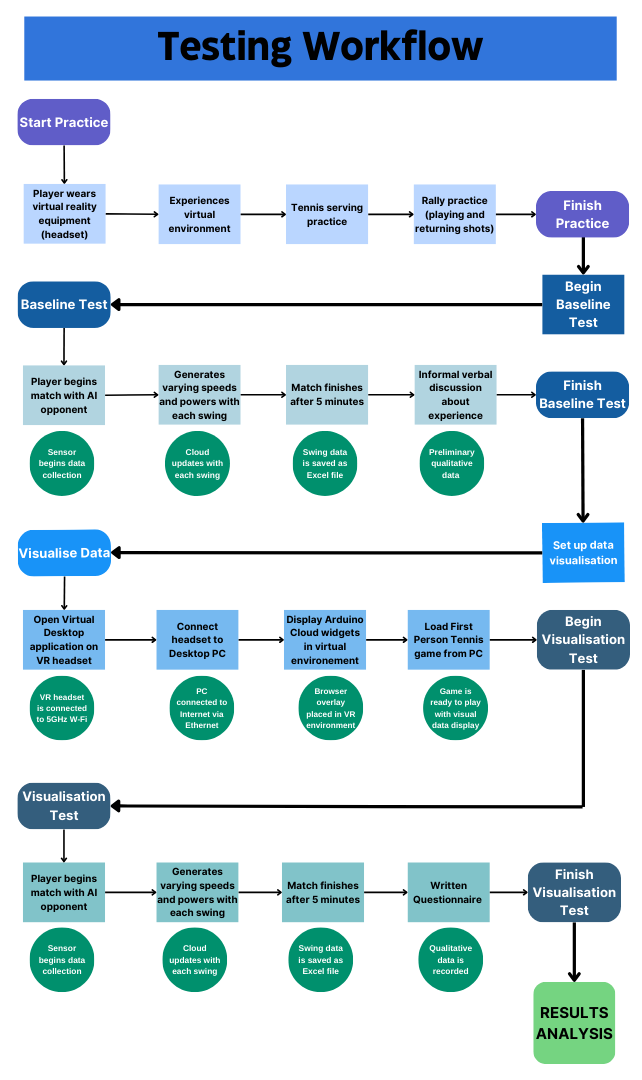}
    \caption{Testing Workflow - A preliminary practice session was followed by the baseline test. After an informal verbal discussion, the participants had a 5-10 minute break while the data visualisation overlay was set up. The visualisation test concluded the physical portion of the study and participants were given a questionnaire after testing.}
    \label{fig4.1}
\end{figure}

\begin{figure}[H]
     \centering
     \includegraphics[scale=0.75]{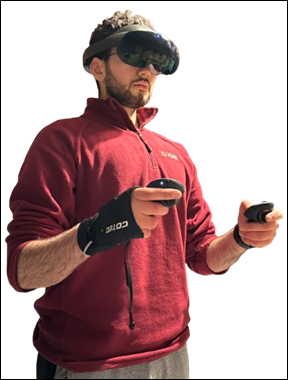}
     \caption{Player with VR headset and glove - Glove is only used on the right hand.}
     \label{fig:Ryan glove}
 \end{figure}

Figure \ref{fig4.1} displays the workflow for a participant for a typical testing process. An initial baseline test was performed in which participants competed against an AI opponent in virtual reality. Quantitative motion data relating to swing speed and swing power were monitored and logged, however the players were not shown these metrics at any point during the game. The data was monitored for a runtime of 5 minutes for each player. Figure \ref{fig:Ryan glove} shows one user holding the VR controllers while wearing the sensor device and VR headset.
\vspace{3mm}

The second test incorporated a visual display of the quantitative data metrics as an overlay in the VR environment. As with the baseline test, participants competed with an AI opponent; however, swing speed and swing power values were shown to the players on the screen in real-time, and so players were able to see swing data as the game was played. The test also concluded after 5 minutes of playing time. 
\vspace{3mm}

Participants were required to complete a qualitative questionnaire after both the baseline and visualisation tests were completed, providing an understanding of the experience of the player.

\subsection{Sensor Integration}

Facilitating wireless data collection from the sensor required an initial wired connection to the PC. The device was calibrated and connected to the Cloud prior to testing, which ensured the Cloud received the correct updates. After completing the wired Cloud connection, the device was removed from the PC and was then powered by the 5 V supply from the LiPo battery and PowerBoost Charger. There was a 2-3 minute delay before updates could be seen in the Cloud Dashboard with the wireless power supply. Swings were emulated with the wireless device before the test begun to ensure new swing readings were made.
\vspace{3mm}

It was important to consider the positioning of the BNO055 Sensor for the duration of the testing phase. In order to obtain the most accurate orientation data (regarding shot angles), the sensor module had to be flush with the player's leading hand. During testing, the hand participants use to hold the controller was constantly changing position and orientation, which may have complicated data collection if not carried out appropriately. By placing the sensor module flat on the back of the player's hand, orientation data was kept consistent with the movement of the hand itself, resulting in an accurate sensor reading.
\vspace{3mm}

The COTEO sleeve \cite{COTEOSleeve} (£14.89) was selected to house the components. The chosen sleeve is a stretch fabric material with two buttoned pouches located in positions: 1. Back of hand, 2. Base of palm (near wrist). This intended function of this product is for use with runners to carry possessions while running, however, it was found to be suitable for housing the device components as it provided space for both the sensor and battery modules in the designated areas of the hand, whilst remaining compact enough to fasten components securely.

\begin{figure}[H]
    \centering
    \includegraphics[scale=1]{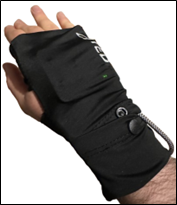}
    \includegraphics[scale=1]{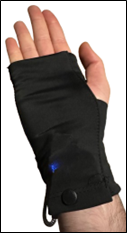}
    \includegraphics[scale=0.9]{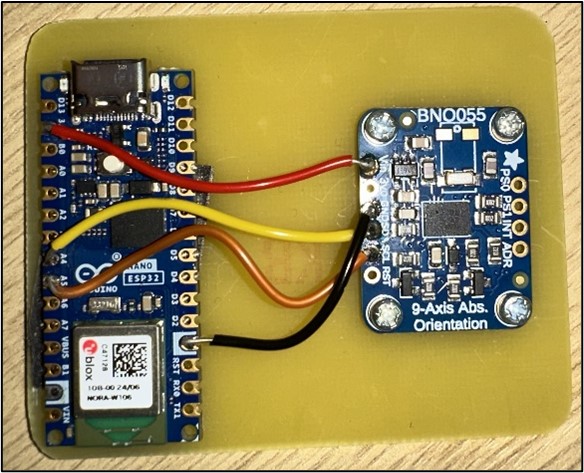}
    \caption{A: Glove Top View (left), B: Bottom View (middle) and C: Nano ESP32 and BNO055 Module - Sensor is placed flat on the back of the users hand and the power supply is positioned on top of the users wrist. Both pouches are sealed with a button.}
    \label{fig:gloves}
\end{figure}

\subsection{Visualising Data}
The initial approach to visualising real-time data in VR tennis involved taking a pre-existing open-source VR tennis game and modifying the source code to include a visual display embedded within the game environment. Two open-source VR tennis games were found.
\vspace{3mm}

Both games are operated with the Unity \cite{Unitycite} gaming development application. This application allows users to develop their own games in 2D or 3D space. Unity offers users complete control over game mechanics and details, meaning visual elements are fully customisable. This initially made the concept of modifying an existing game to integrate a visual data display more attractive. The problems associated with this method involved poor game functionality. The two existing games required further modifications to run successfully. This would have delayed progress greatly on the project which is why this method was not chosen.
\vspace{3mm}

The final approach uses an overlay of swing data layered on top of the VR environment, rather than a modification to an existing game. The complete data visualisation process can be seen in Figure 8 and an example of the visualisation within the virtual tennis environment is shown in Figure 9.
\vspace{3mm}

Figure 7 shows the visualised swing data in the Arduino Cloud Dashboard. Swing speed and swing power are declared as floating-point numbers in the Cloud and are linked to visual widgets in the Cloud Dashboard. The widgets for both metrics are represented by gauges, which displays a numerical value corresponding to each swing. 
\vspace{3mm}

The Cloud Dashboard is accessible using an Internet web browser. To visually show the widgets to players in the VR environment, the web browser must be made available for display using the VR headset. There are requirements that must be met to ensure the display is viable:

\begin{enumerate}
    \item The VR tennis game must be PC-VR compatible.
    \item The desktop computer used must have a compatible GPU to run the tennis game.
    \item The computer must be connected to a 5 GHz Wi-Fi channel using an Ethernet cable.
\end{enumerate}

As swing speed and swing power were visualised in the cloud Dashboard, which was accessible via internet web browser, the browser had to be seen in the virtual environment whilst the VR tennis game was being played.

\begin{center}
    \begin{figure}[h!]
        \centering
        \fbox{\includegraphics[scale=0.22]{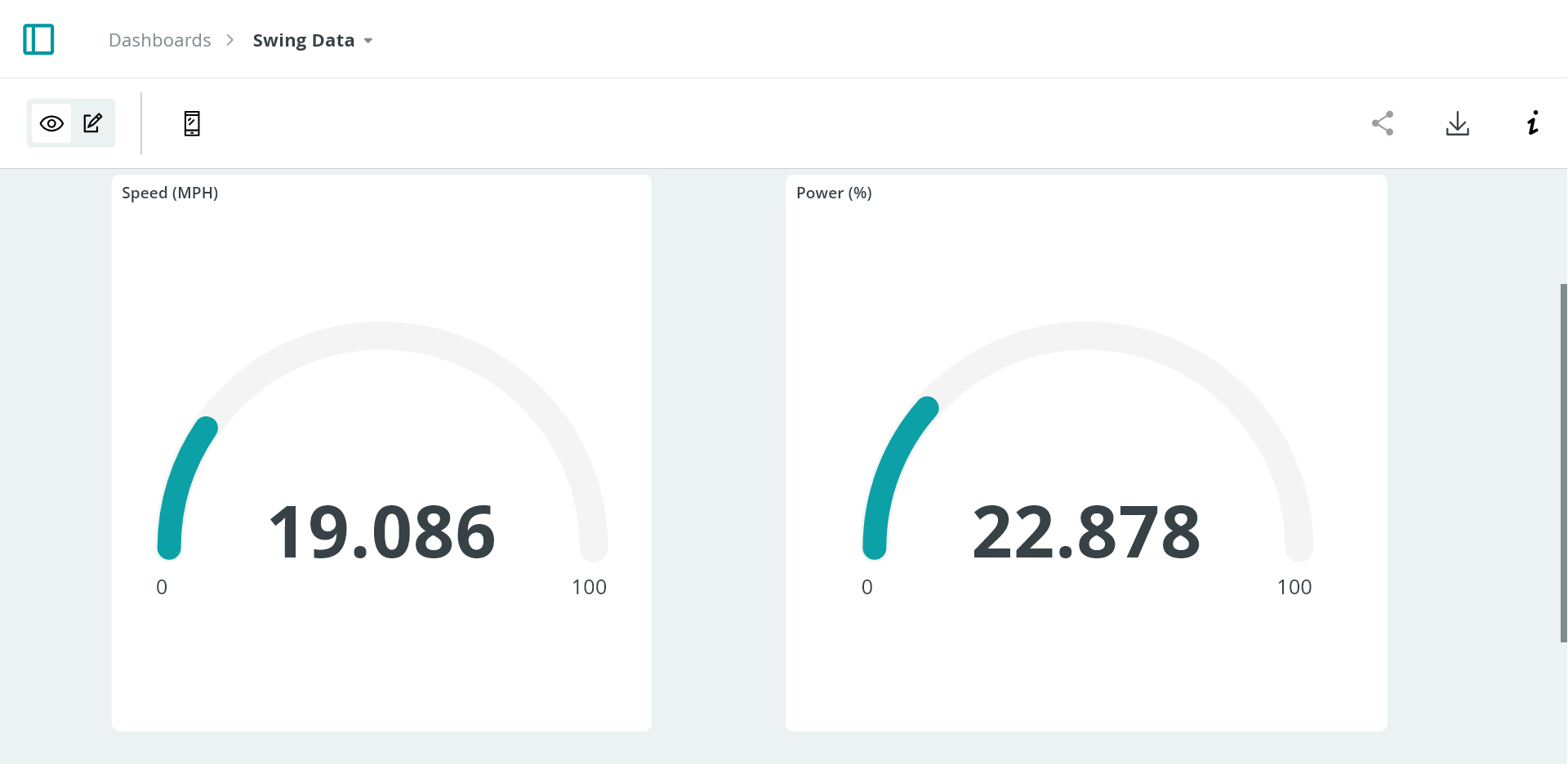}}
        \caption{Arduino Cloud Dashboard with visual display widgets - Both widgets are gauges that numerically display the respective value of swing speed in mph (left) and swing power as a percentage (right).}
        \label{fig:Widget display}
    \end{figure}
\end{center}

\vspace{3mm}

\textbf{Virtual Desktop Interface}: Virtual Desktop \cite{Virtualdesktop} is a VR software application. This enables a wireless connection between the VR headset and PC to be made. Both the headset and PC had to be connected to the same Internet network to allow the connection to be made. Both connections were required to be on a 5 GHz channel.
\vspace{3mm}

Once both the headset and PC were connected to the same network, the desktop was made available for wireless connection in the Virtual Desktop app within the headset. Connection problems arose if the PC is connected to the Internet using a wireless connection (using a Wi-Fi card or dongle). If the PC was connected using a wireless Wi-Fi connection, the Virtual Desktop application would not run at a proficient level, resulting in a slow experience. Therefore, the PC was connected to the Internet using an Ethernet cable, providing a quality experience.
\vspace{3mm}

Usually, it is not possible to run applications simultaneously on the VR headset, therefore the internet browser application for the Meta Quest Pro could not be viewed whilst playing First Person Tennis from the headset itself. 
Virtual Desktop ensures that VR games can be played using VR equipment, but must run from the connected PC. Therefore, the VR game must be PC-VR compatible and the PC must meet the minimum requirements to run the tennis game, hence First Person Tennis is used as it is PC-VR compatible. The game offers users an automatic teleportation option to simulate court movement, which was chosen due to spatial limitations. The minimum desktop requirements to run First Person Tennis are described as stated by Steam \cite{steamFPT}:
\begin{enumerate}
    \item 64-bit processor and operating system
    \item Operating System: Windows 10/8/7/Vista/XP.
    \item Processor: 2.66 GHz Intel Core 2 Quad Q8400, 3.0 GHz AMD Phenom II X4 940.
    \item Memory: 4 GB RAM.
    \item Graphics: NVIDIA GTX 1060 / RADEON RX 580.
    \item DirectX: Version 9.0c.
    \item Storage: 3 GB available space.
    \item VR Support: SteamVR
\end{enumerate}

\begin{figure}[H]
    \centering
    \includegraphics[scale = 0.54]{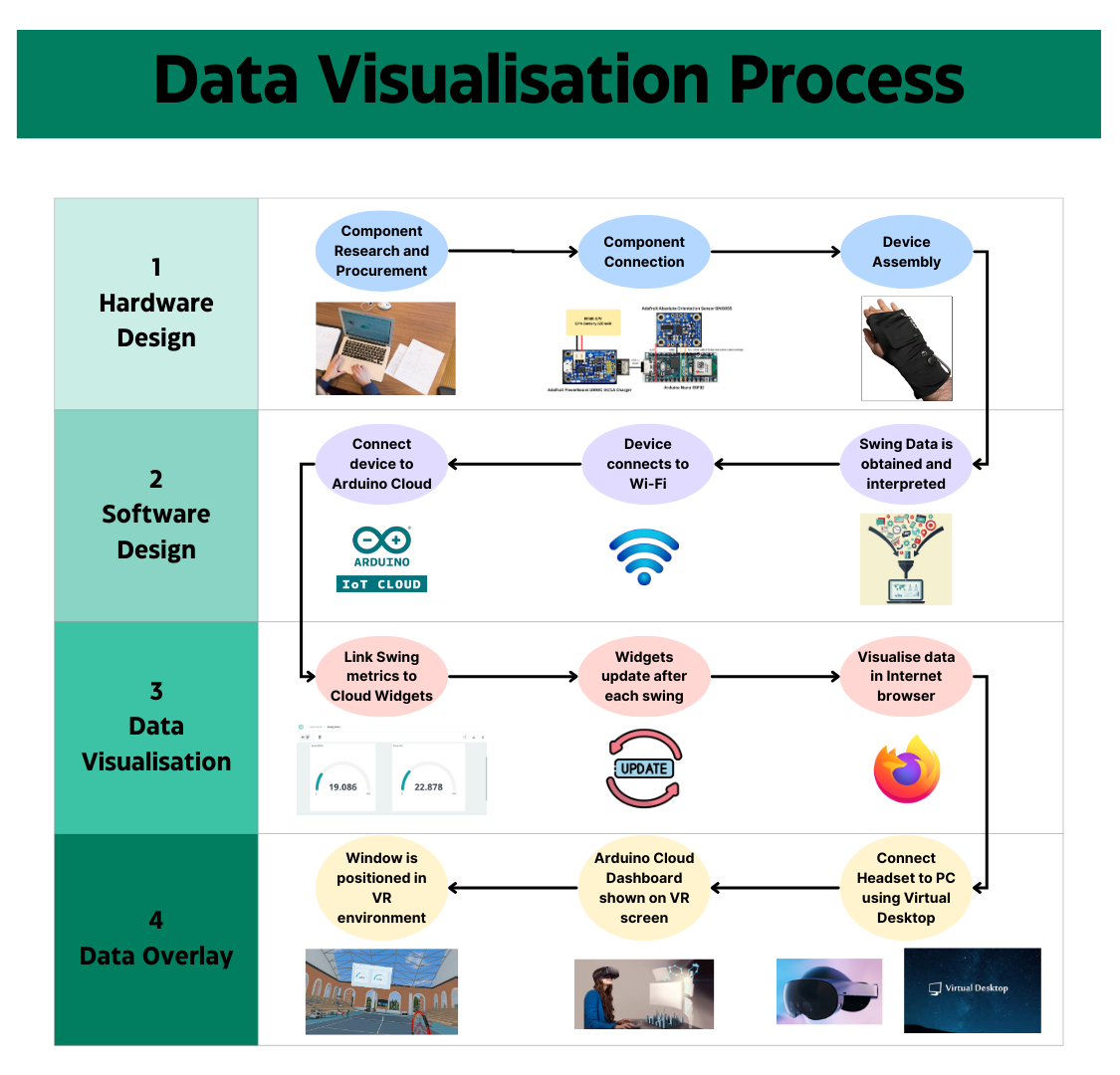}
    \caption{Data Visualisation Process - Steps outlining the process to facilitate data visualisation: 1. Hardware Design - Physical prototyping and assembly, 2. Software Design - Wireless connectivity and Cloud configuration, 3. Data Visualisation - Real-time metric updates, and 4. Data Overlay - Visual window display in VR.}
    \label{fig4.6}
\end{figure}

\subsubsection{Participant Interaction}
Participants were given control of positioning the floating window overlay in the virtual environment.
Some participants chose to position the window above the court so that it was still in their field of vision while playing. Others positioned the window to the side of the court so that they would have an unobstructed view and would have to turn their head to see the data.

\begin{figure}[H]
    \centering
    \includegraphics[scale = 0.12]{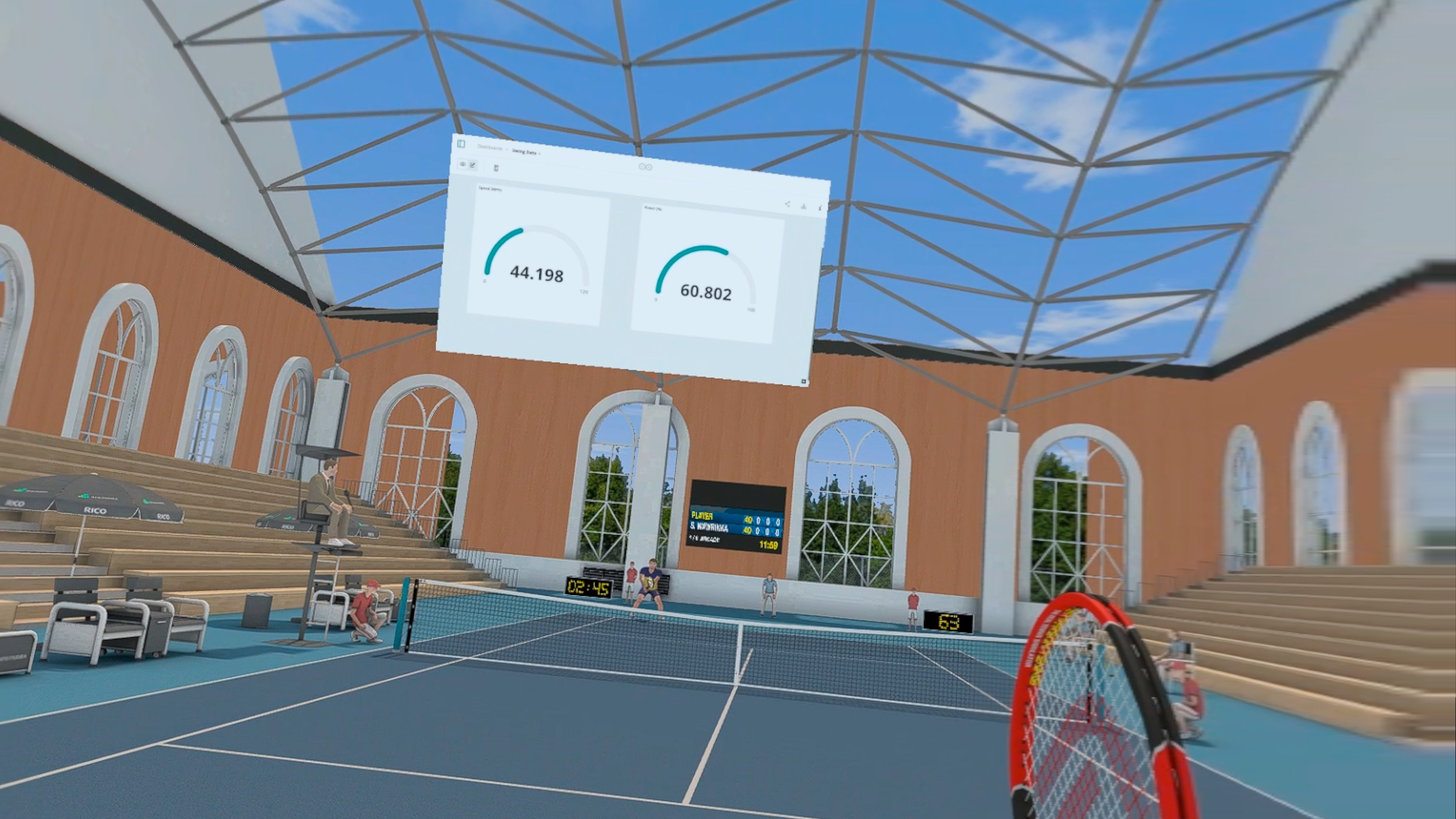}
    \caption{Visualisation window floating in-game - This example shows the window positioned slightly above the court, which allows players to view the data metrics without having to look to the side. \textit{N.B. The swing data window can be moved to an optimum position on screen by the player}.}
    \label{fig:VR overlay}
\end{figure}

\subsection{Questionnaire}

All participants were provided a questionnaire with questions of varying styles, (descriptive and multiple choice), and were asked to gauge the general feeling of the participants in order to identify any positive elements and any future improvements. The questions given to the participants were inspired from previously validated instruments, which are readily available in the literature \cite{Wu2021SPinPong}.

\section{Results}
The data collected from both tests was analysed to compare performance under both conditions. A comparison is made to evaluate the change in performance level with due consideration of the swing speed and shot count between both conditions. In the evaluation of shot accuracy, an "accurate shot" was considered a shot in which the ball had landed in a playable area of the tennis court. A ``point scored'' was defined as the result of playing a valid shot that won the point. Quantitative data relating to swing power was also analysed.

\subsection{Quantitative Results}
Figures 10 to 15 show the percentage of total shots and accurate shots made at varying swing speeds in both tests.

\begin{figure}[H]
    \centering
    \includegraphics[scale=1]{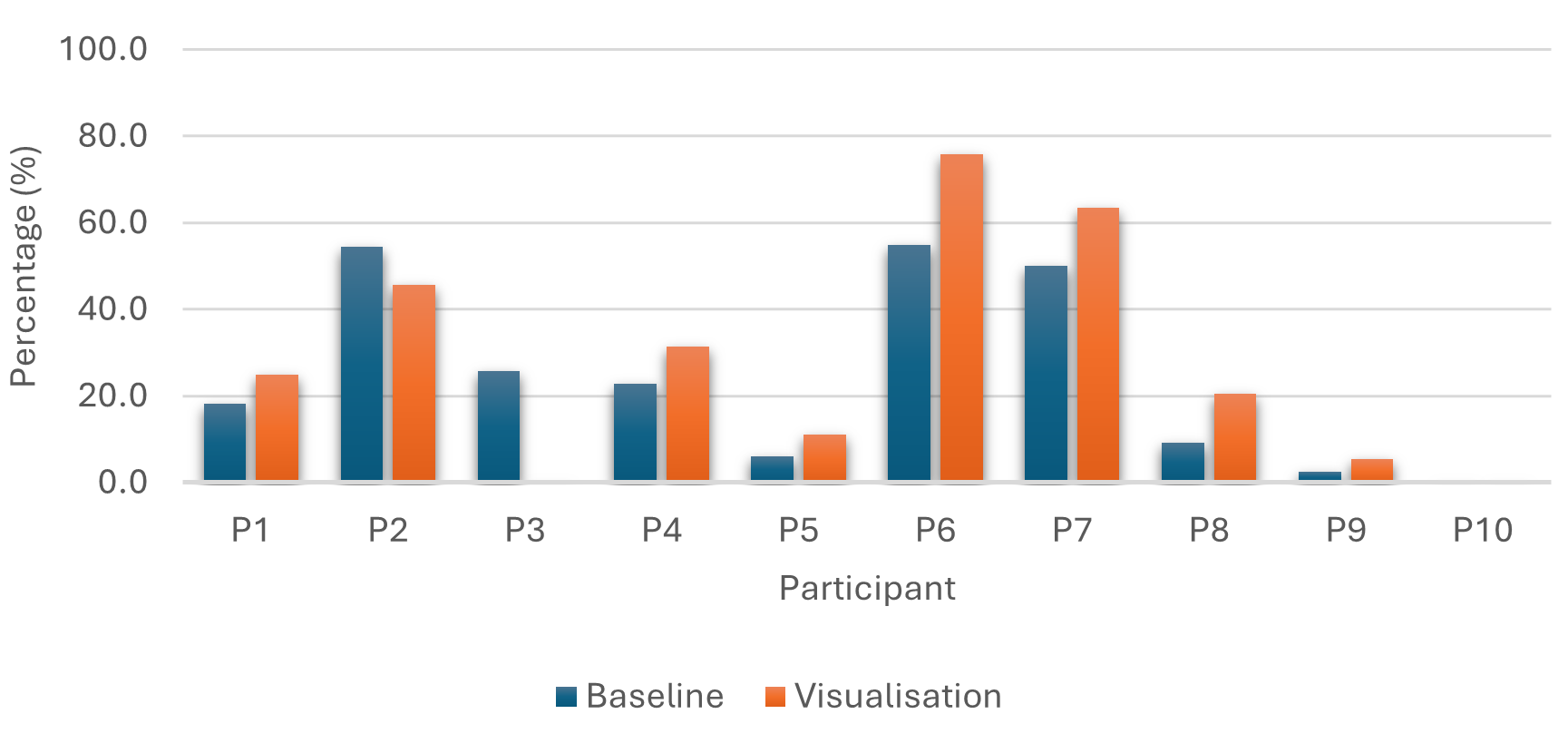}
    \caption{Percentage of Total Shots for swing speeds greater than 40 mph in both tests. Blue indicates results under the baseline condition and orange indicates results under the visualisation condition.}
    \label{fig:10}
\end{figure}

\begin{figure}[H]
    \centering
    \includegraphics[scale=1]{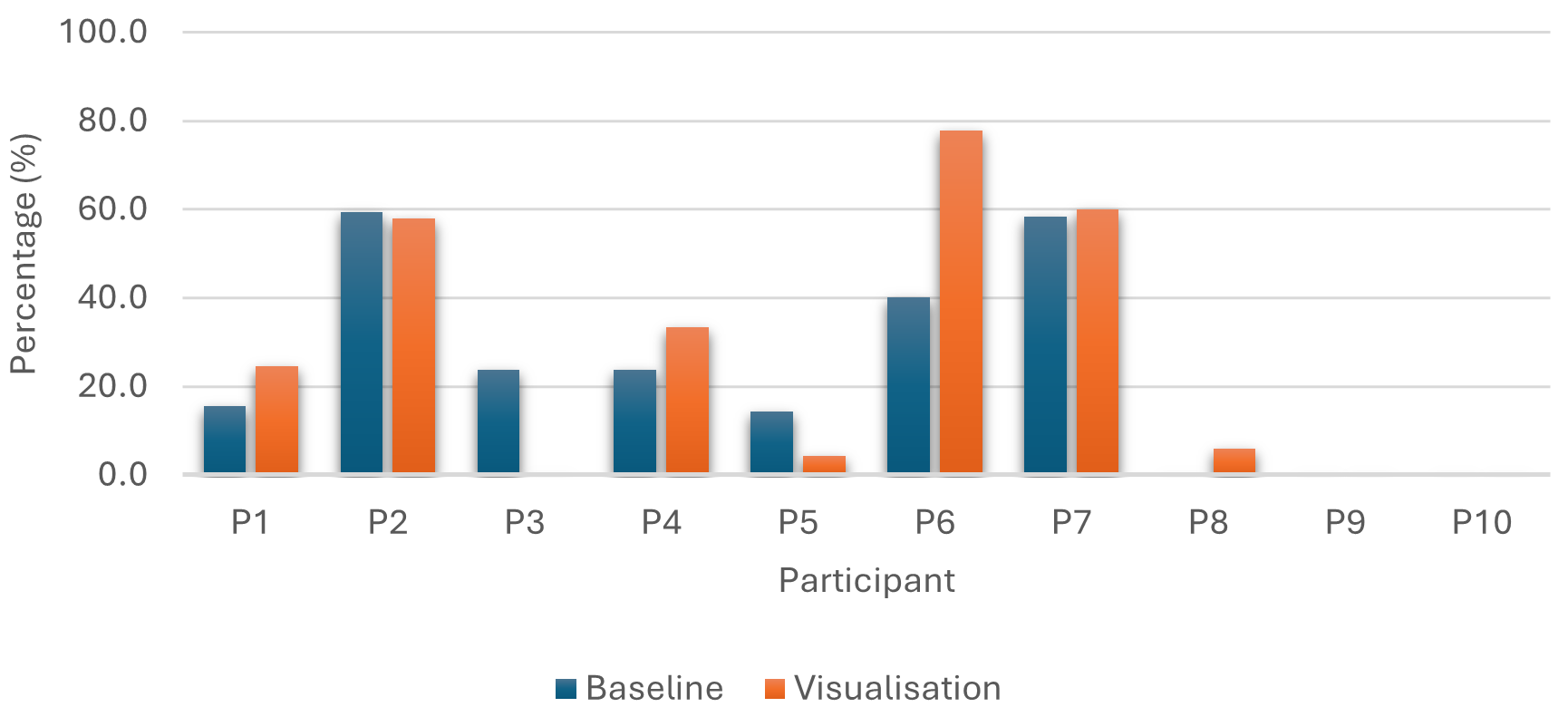}
    \caption{Percentage of accurate shots for swing speeds greater than 40 mph in both tests. Blue indicates results under the baseline condition and orange indicates results under the visualisation condition.}
    \label{fig:11}
\end{figure}

\begin{figure}[H]
    \centering
    \includegraphics[scale=1]{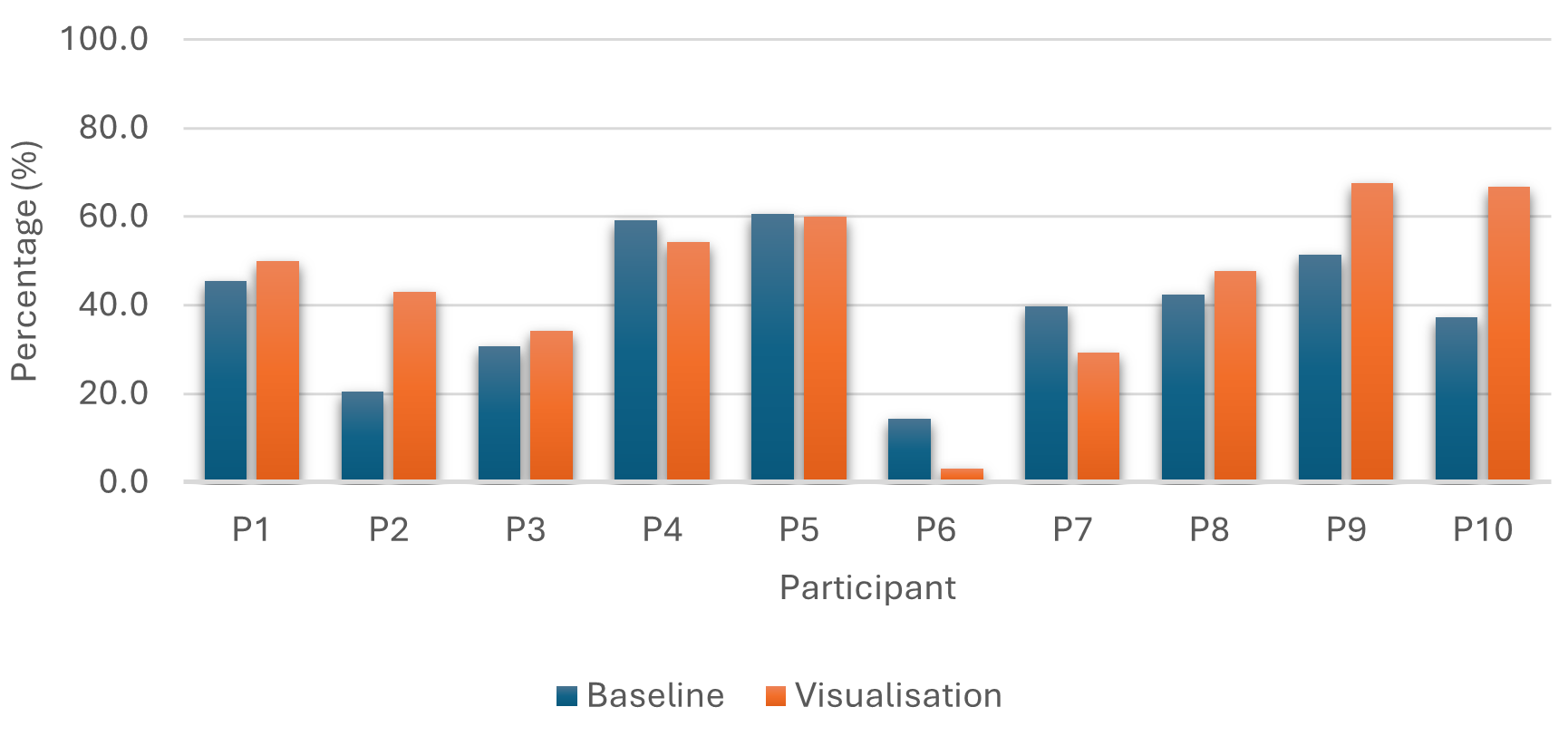}
    \caption{Percentage of total shots for swing speeds between 25 mph and 40 mph in both tests. Blue indicates results under the baseline condition and orange indicates results under the visualisation condition.}
    \label{fig:12}
\end{figure}

\begin{figure}[H]
    \centering
    \includegraphics[scale=1]{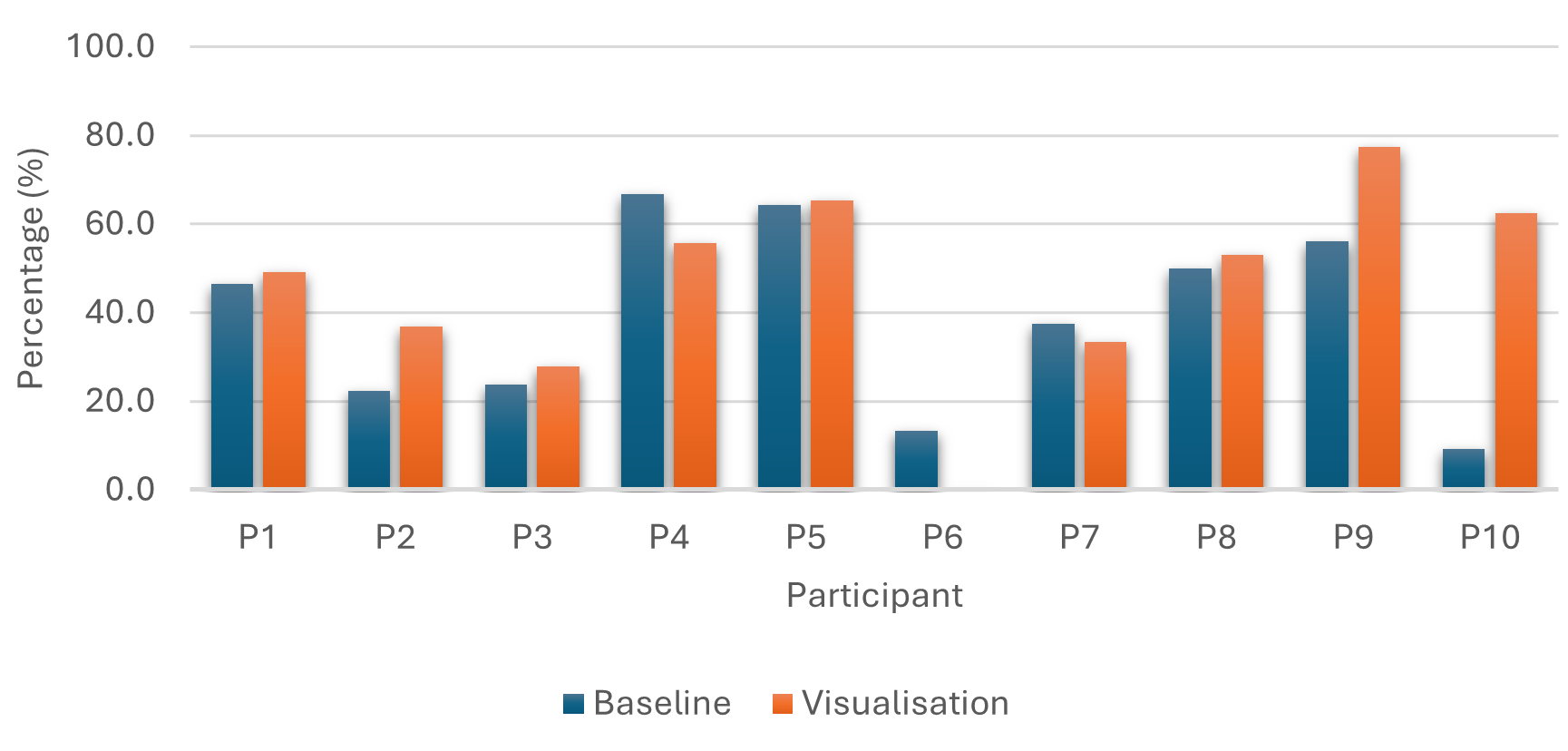}
    \caption{Percentage of accurate shots for swing speeds greater than 40 mph in both tests. Blue indicates results under the baseline condition and orange indicates results under the visualisation condition.}
    \label{fig:13}
\end{figure}

\begin{figure}[H]
    \centering
    \includegraphics[scale=1]{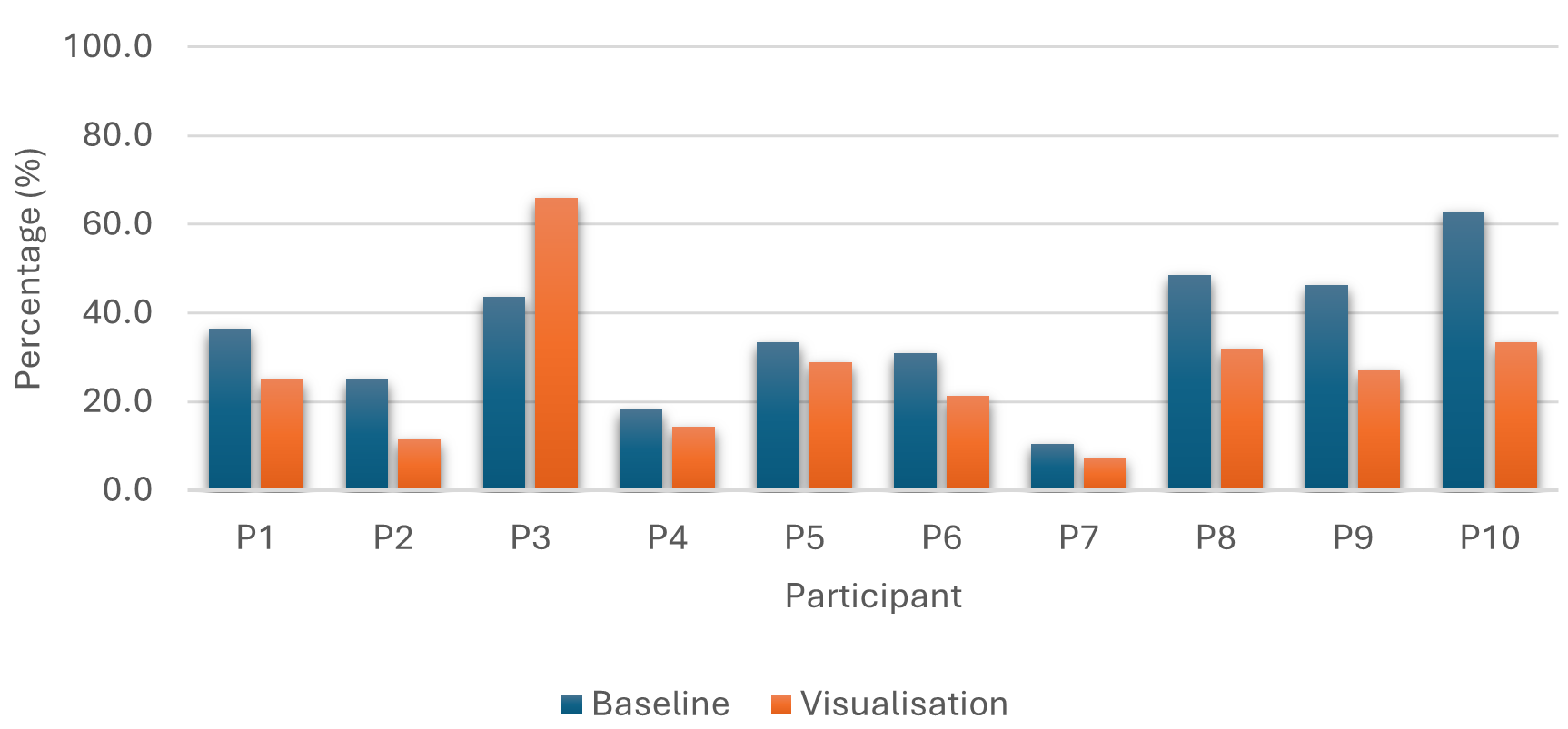}
    \caption{Percentage of total shots for swing speeds less than 25 mph in both tests. Blue indicates results under the baseline condition and orange indicates results under the visualisation condition.}
    \label{fig:14}
\end{figure}

\begin{figure}[H]
    \centering
    \includegraphics[scale=1]{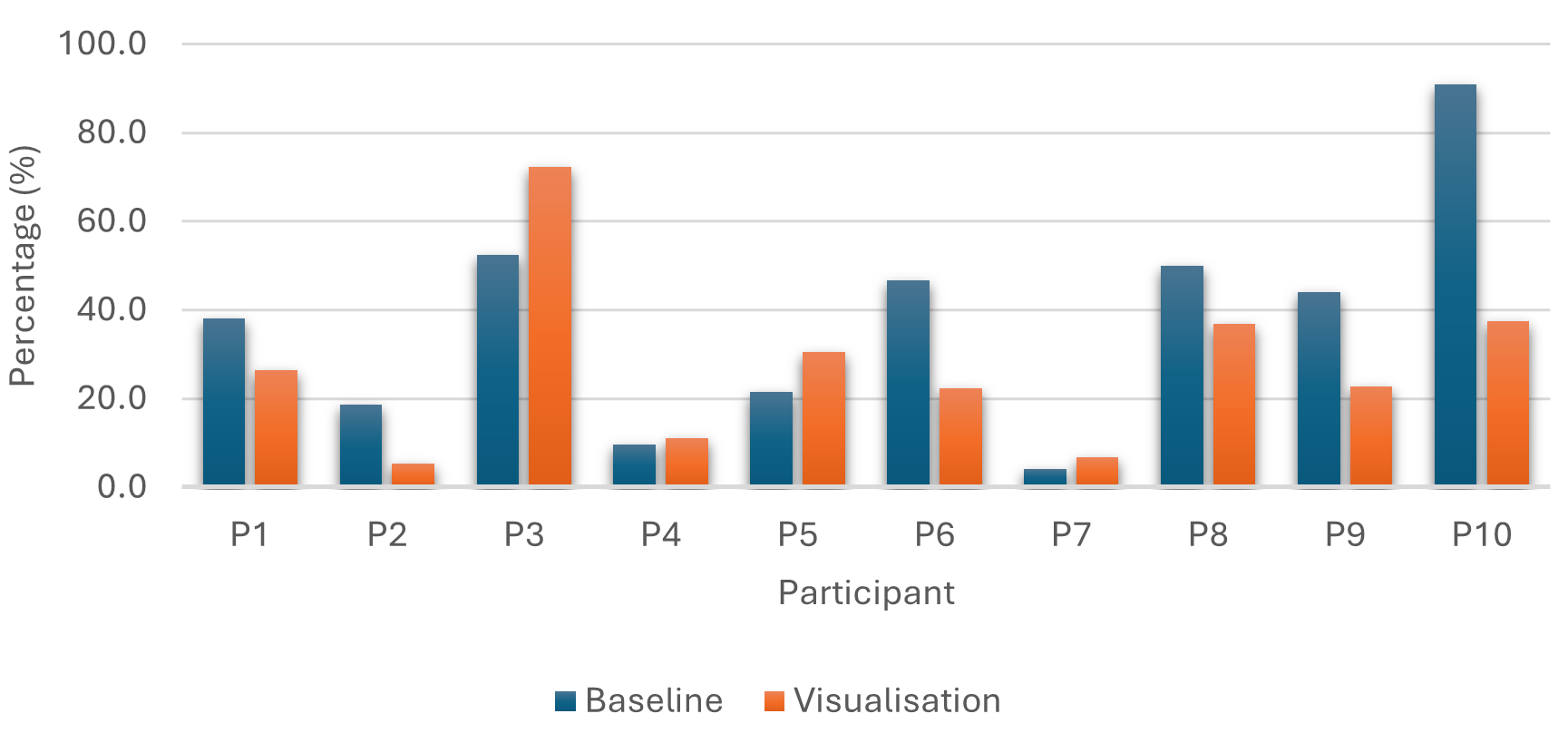}
    \caption{Percentage of accurate shots for swing speeds less than 25 mph in both tests. Blue indicates results under the baseline condition and orange indicates results under the visualisation condition.}
    \label{fig:15}
\end{figure}

Tables 1 to 7 contain the numerical data obtained from the study. Table 1 summarises the change in the percentage of accurate shots between both conditions. Six out of the ten participants had shown to decrease the percentage of accurate shots made when switching to the second condition, with a decrease in 22\% for Participant 4 being the greatest change. The remaining four participants had shown an increase in accurate shot percentage, with Participant 8 achieving the highest improvement of 21\% more accurate shots.
\vspace{3mm}

Table 2 displays a comparison of the number of points won in both conditions. After 5 minutes of playing time against the AI opponent, six participants played more winning shots in the second test, with two participants unchanged and two with a worse score.

\begin{table}[H]
\centering
\begin{tabular}{ |>{\centering\arraybackslash}p{3.5cm}|>{\centering\arraybackslash}p{3.5cm}|>{\centering\arraybackslash}p{3.5cm}|>{\centering\arraybackslash}p{3.5cm}| }
 \hline
 \multicolumn{4}{|c|}{\cellcolor{lightgray}\large Percentage of Accurate Shots in both conditions} \\
 \hline
 \bf{Participant}& \bf{Baseline Accurate Shots (\%)} &\bf{Visualisation Accurate Shots(\%)}&\bf{Change in Accurate Shots(\%)}\\
 \hline
 Participant 1 & 92\% & 83\% & \cellcolor{red}-9\% \\
 \hline
 Participant 2 & 61\% & 54\% & \cellcolor{red}-7\% \\
 \hline
 Participant 3 & 35\% & 47\% & \cellcolor{green}+12\% \\
 \hline
 Participant 4 & 48\% & 26\% & \cellcolor{red}-22\% \\
 \hline
 Participant 5 & 42\% & 51\% & \cellcolor{green}+9\% \\
 \hline
 Participant 6 & 36\% & 27\% & \cellcolor{red}-9\% \\
 \hline
 Participant 7 & 50\% & 37\% & \cellcolor{red}-13\% \\
 \hline
 Participant 8 & 18\% & 39\% & \cellcolor{green}+21\% \\
 \hline
 Participant 9 & 64\% & 59\% & \cellcolor{red}-5\% \\
 \hline
 Participant 10 & 31\% & 48\% & \cellcolor{green}+17\% \\
 \hline
\end{tabular}
\caption{Table of Percentage of Accurate Shots Comparison under both conditions - Red indicates a worse performance and green indicates a better performance. 6/10 participants achieved a lower percentage of accurate shots in the visualisation condition.}
\label{fig:Table1}
\end{table}

\begin{table}[H]
\centering
\begin{tabular}{ |>{\centering\arraybackslash}p{3.5cm}|>{\centering\arraybackslash}p{3.5cm}|>{\centering\arraybackslash}p{3.5cm}|>{\centering\arraybackslash}p{3.5cm}| }
 \hline
 \multicolumn{4}{|c|}{\cellcolor{lightgray}\large Number of Points Won in both conditions} \\
 \hline
 \bf{Participant}& \bf{Baseline Points Scored} &\bf{Visualisation Points Scored}&\bf{Change in Points Scored}\\
 \hline
 Participant 1& 6    &11&   \cellcolor{green}+5\\
 \hline
 Participant 2&  5  & 4   &\cellcolor{red}-1\\
 \hline
 Participant 3& 2& 4&  \cellcolor{green}+2\\
 \hline
 Participant 4& 1 & 1 &  \cellcolor{gray}0\\
 \hline
 Participant 5& 2  & 6& \cellcolor{green}+4\\
 \hline
 Participant 6& 1  & 1   & \cellcolor{gray}0\\
 \hline
 Participant 7& 2  & 3 & \cellcolor{green}+1\\
 \hline
 Participant 8& 0  & 1 & \cellcolor{green}+1\\
 \hline
 Participant 9& 4  & 6 & \cellcolor{green}+2\\
 \hline
 Participant 10& 6  & 3 & \cellcolor{red}-3\\
 \hline
 \textbf{Mean}&\cellcolor{lightgray} 2.9&\cellcolor{lightgray}4\%&\cellcolor{lightgray}1.1\\
 \hline
\end{tabular}
\caption{Table of Number of Points won comparison in both conditions - Red indicates the participants with less points won and green indicates the participants with more points won. Grey indicates the participants with an unchanged number of points won. 6/10 participants played more shots that won a point in the visualisation condition. }
\label{fig:Table2}
\end{table}
\vspace{3mm}

\begin{table}[H]
\centering
\begin{tabular}{ |>{\centering\arraybackslash}p{3.5cm}|>{\centering\arraybackslash}p{3.5cm}|>{\centering\arraybackslash}p{3.5cm}|>{\centering\arraybackslash}p{3.5cm}| }
 \hline
 \multicolumn{4}{|c|}{\cellcolor{lightgray}\large Statistical Analysis for Number of Points won} \\
 \hline
 \bf{Metric}& \bf{Mean Difference (V - B)} &\bf{t-statistic}&\bf{p-value}\\
 \hline
 No. of points won & 1.1 & 1.4923 & 0.1698\\
 \hline
\end{tabular}
\caption{Statistical analysis for accurate shots played at varying swing speeds.}
\label{fig:Table3}
\end{table}

Tables 4 to 8 display shot data at varying swing speeds for both conditions. Tables 3 and 4 display the percentage of total shots taken for three swing speed brackets for each participant. Looking at Participant 1 for example, 18.2\% of their total shots were at a swing speed greater than 40 mph in the baseline condition, however in the visualisation condition they had a higher percentage of shots at swing speeds greater than 40 mph, at 25\%. 
\newline
The cells are highlighted in green or red to indicate a higher (green) or lower (red) percentage between both conditions.
\vspace{3mm}

7/10 Participants increased the number of shots taken at swing speeds greater than 40 mph, 6/10 at swing speeds between 25 mph and 40 mph, and only 2/10 at swing speeds less than 25 mph.
\vspace{3mm}

This shows that the majority of players were aiming to generate swing speeds higher than 40 mph more frequently after seeing their metrics visualised in the visualisation condition.

\begin{table}[H]
\centering
\begin{tabular}{ |>{\centering\arraybackslash}p{3.5cm}|>{\centering\arraybackslash}p{3.5cm}|>{\centering\arraybackslash}p{3.5cm}|>{\centering\arraybackslash}p{3.5cm}| }
 \hline
 \multicolumn{4}{|c|}{\cellcolor{lightgray}\large Total Shots at different swing speeds (BASELINE)} \\
 \hline
 \bf{Participant} & \bf{Greater than 40 mph} & \bf{Between 25 mph and 40 mph} & \bf{Less than 25 mph} \\
 \hline
  Participant 1 & \cellcolor{red}18.2\% &\cellcolor{red} 45.5\% & \cellcolor{green}36.4\% \\
 \hline
 Participant 2 & \cellcolor{green}54.5\% & \cellcolor{red}20.5\% & \cellcolor{green}25.0\% \\
 \hline
 Participant 3 & \cellcolor{green}25.8\% & \cellcolor{red}30.6\% & \cellcolor{red}43.5\% \\
 \hline
 Participant 4 & \cellcolor{red}22.7\% & \cellcolor{green}59.1\% & \cellcolor{green}18.2\% \\
 \hline
 Participant 5 & \cellcolor{red}6.1\% & \cellcolor{green}60.6\% & \cellcolor{green}33.3\% \\
 \hline
 Participant 6 & \cellcolor{red}54.8\% & \cellcolor{green}14.3\% & \cellcolor{red}10.4\% \\
 \hline
 Participant 7 & \cellcolor{red}50.0\% & \cellcolor{green}39.6\% & \cellcolor{green}10.4\% \\
 \hline
 Participant 8 & \cellcolor{red}9.1\% & \cellcolor{red}42.4\% & \cellcolor{green}48.5\% \\
 \hline
 Participant 9 & \cellcolor{red}2.6\% & \cellcolor{red}51.3\% & \cellcolor{green}46.2\% \\
 \hline
 Participant 10 & \cellcolor{gray}0.0\% & \cellcolor{red}37.1\% & \cellcolor{green}62.9\% \\
 \hline
 \textbf{Mean}&\cellcolor{lightgray} 24.4\%&\cellcolor{lightgray}40.1\%&\cellcolor{lightgray}35.5\%\\
 \hline
\end{tabular}
\caption{Swing Speed Proportions for the Baseline Test, showing the percentage of total shots with swings at varying speeds ($>$40 mph, 25 mph$<$$>$40 mph and $<$25 mph). Red indicates a worse performance and green indicates a better performance under this condition. Grey indicates an unchanged performance. 8/10 participants played more shots at swing speeds lower than 25 mph.}
\label{fig:Table4}
\end{table}

\begin{table}[H]
\centering
\begin{tabular}{ |>{\centering\arraybackslash}p{3.5cm}|>{\centering\arraybackslash}p{3.5cm}|>{\centering\arraybackslash}p{3.5cm}|>{\centering\arraybackslash}p{3.5cm}| }
 \hline
 \multicolumn{4}{|c|}{\cellcolor{lightgray}\large Total Shots at different swing speeds (VISUALISATION)} \\
 \hline
 \bf{Participant} & \bf{Greater than 40 mph} & \bf{Between 25 mph and 40 mph} & \bf{Less than 25 mph} \\
 \hline
 Participant 1 & \cellcolor{green}25.0\% & \cellcolor{green}50.0\% & \cellcolor{red}25.0\% \\
 \hline
 Participant 2 & \cellcolor{red}45.7\% & \cellcolor{green}42.9\% & \cellcolor{red}11.4\% \\
 \hline
 Participant 3 & \cellcolor{red}0.0\% & \cellcolor{green}34.2\% & \cellcolor{green}65.8\% \\
 \hline
 Participant 4 & \cellcolor{green}31.4\% & \cellcolor{red}54.3\% & \cellcolor{red}14.3\% \\
 \hline
 Participant 5 & \cellcolor{green}11.1\% & \cellcolor{red}60.0\% & \cellcolor{red}28.9\% \\
 \hline
 Participant 6 & \cellcolor{green}75.8\% & \cellcolor{red}3.0\% & \cellcolor{green}21.2\% \\
 \hline
 Participant 7 & \cellcolor{green}63.4\% & \cellcolor{red}29.3\% & \cellcolor{red}7.3\% \\
 \hline
 Participant 8 & \cellcolor{green}20.5\% & \cellcolor{green}47.7\% & \cellcolor{red}31.8\% \\
 \hline
 Participant 9 & \cellcolor{green}5.4\% & \cellcolor{green}67.6\% & \cellcolor{red}27.0\% \\
 \hline
 Participant 10 & \cellcolor{gray}0.0\% & \cellcolor{green}66.7\% & \cellcolor{red}33.3\% \\
 \hline
 \textbf{Mean}&\cellcolor{lightgray} 27.8\%&\cellcolor{lightgray}45.6\%&\cellcolor{lightgray}26.6\%\\
 \hline
\end{tabular}
\caption{Swing Speed Proportions for the Visualisation Test, showing the percentage of swings at varying speeds ($>$40 mph, 25 mph$<$$>$40 mph and $<$25 mph). Red indicates a worse performance and green indicates a better performance under this condition. Grey indicates an unchanged performance. 7/10 participants played more shots at swing speeds above 40 mph, and 6/10 played more shots at swing speeds between 25 mph and 40 mph under the visualisation condition. 8/10 participants played less shots at swing speeds less than 25 mph under the visualisation condition.}
\label{fig:Table5}
\end{table}
\vspace{3mm}

\begin{table}[H]
\centering
\begin{tabular}{ |>{\centering\arraybackslash}p{3.5cm}|>{\centering\arraybackslash}p{3.5cm}|>{\centering\arraybackslash}p{3.5cm}|>{\centering\arraybackslash}p{3.5cm}| }
 \hline
 \multicolumn{4}{|c|}{\cellcolor{lightgray}\large Statistical Analysis for Total Shots at varying swing speeds} \\
 \hline
 \bf{Participant}& \bf{Mean Difference (V - B)} &\bf{t-statistic}&\bf{p-value}\\
 \hline
 Greater than 40 mph & 3.5\% & 0.8378 & 0.4238\\
 \hline
 Between 25 mph and 40 mph & 5.5\% & 1.2731 & 0.2349 \\
 \hline
 Less than 25 mph & -8.9\% & -2.0723 & N/A \\
 \hline
\end{tabular}
\caption{Statistical analysis for accurate shots played at varying swing speeds.}
\label{fig:Table6}
\end{table}

Tables 6 and 7 display the percentage of Accurate Shots played at varying swing speeds for both conditions. Comparing the two sets of results, there was a significant increase in the number of participants with higher accuracies in the 25 mph to 40 mph range for the visualisation condition (3/10 baseline to 7/10 visualisation).
Additionally, the results have shown that more participants increased the percentage of accurate shots at swing speeds greater than 40 mph in the visualisation condition (3/10 baseline to 5/10 visualisation). 

\vspace{3mm}

These results clearly indicate the influence of the visualised data metrics in the visualisation condition.

\begin{table}[H]
\centering
\begin{tabular}{ |>{\centering\arraybackslash}p{3.5cm}|>{\centering\arraybackslash}p{3.5cm}|>{\centering\arraybackslash}p{3.5cm}|>{\centering\arraybackslash}p{3.5cm}| }
 \hline
 \multicolumn{4}{|c|}{\cellcolor{lightgray}\large Accurate Shots at different swing speeds (BASELINE)} \\
 \hline
 \bf{Participant} & \bf{Greater than 40 mph} & \bf{Between 25 mph and 40 mph} & \bf{Less than 25 mph} \\
 \hline
 Participant 1 & \cellcolor{red}15.5\% & \cellcolor{red}46.5\% & \cellcolor{green}38.0\% \\
\hline
 Participant 2 & \cellcolor{green}59.3\% & \cellcolor{red}22.2\% & \cellcolor{green}18.5\% \\
\hline
 Participant 3 & \cellcolor{green}23.8\% & \cellcolor{red}23.8\% & \cellcolor{red}52.4\% \\
\hline
 Participant 4 & \cellcolor{red}23.8\% & \cellcolor{green}66.7\% & \cellcolor{red}9.5\% \\
\hline
 Participant 5 & \cellcolor{green}14.3\% & \cellcolor{red}64.3\% & \cellcolor{red}21.4\% \\
\hline
 Participant 6 & \cellcolor{red}40.0\% & \cellcolor{green}13.3\% & \cellcolor{red}46.7\% \\
\hline
 Participant 7 & \cellcolor{red}58.3\% & \cellcolor{green}37.5\% & \cellcolor{red}4.2\% \\
\hline
 Participant 8 & \cellcolor{red}0.0\% & \cellcolor{red}50.0\% & \cellcolor{green}50.0\% \\
\hline
 Participant 9 & \cellcolor{gray}0.0\% & \cellcolor{red}56.0\% & \cellcolor{green}44.0\% \\
\hline
 Participant 10 & \cellcolor{gray}0.0\% & \cellcolor{red}9.1\% & \cellcolor{green}90.9\% \\
 \hline
 \textbf{Mean}&\cellcolor{lightgray} 23.5\%&\cellcolor{lightgray}38.9\%&\cellcolor{lightgray}37.6\%\\
 \hline
\end{tabular}
\caption{Accurate Shot percentages for the Baseline Test, showing the percentage of accurate shots for various swing speeds ($>$40 mph, 25 mph$<$$>$40 mph and $<$25 mph). Red indicates a worse performance and green indicates a better performance. Grey indicates no change in performance.}
\label{fig:Table7}
\end{table}

\begin{table}[H]
\centering
\begin{tabular}{ |>{\centering\arraybackslash}p{3.5cm}|>{\centering\arraybackslash}p{3.5cm}|>{\centering\arraybackslash}p{3.5cm}|>{\centering\arraybackslash}p{3.5cm}| }
 \hline
 \multicolumn{4}{|c|}{\cellcolor{lightgray}\large Accurate Shots at different swing speeds (VISUALISATION)} \\
 \hline
 \bf{Participant} & \bf{Greater than 40 mph} & \bf{Between 25 mph and 40 mph} & \bf{Less than 25 mph} \\
 \hline
 Participant 1 & \cellcolor{green}24.5\% & \cellcolor{green}49.1\% & \cellcolor{red}26.4\% \\
\hline
 Participant 2 & \cellcolor{red}57.9\% & \cellcolor{green}36.8\% & \cellcolor{red}5.3\% \\
\hline
 Participant 3 & \cellcolor{red}0.0\% & \cellcolor{green}27.8\% & \cellcolor{green}72.2\% \\
\hline
 Participant 4 & \cellcolor{green}33.3\% & \cellcolor{red}55.6\% & \cellcolor{green}11.1\% \\
\hline
 Participant 5 & \cellcolor{red}4.3\% & \cellcolor{green}65.2\% & \cellcolor{green}30.4\% \\
\hline
 Participant 6 & \cellcolor{green}77.8\% & \cellcolor{red}0.0\% & \cellcolor{green}22.2\% \\
\hline
 Participant 7 & \cellcolor{green}60.0\% & \cellcolor{red}33.3\% & \cellcolor{green}6.7\% \\
\hline
 Participant 8 & \cellcolor{green}5.9\% & \cellcolor{green}52.9\% & \cellcolor{red}36.8\% \\
\hline
 Participant 9 & \cellcolor{gray}0.0\% & \cellcolor{green}77.3\% & \cellcolor{red}22.7\% \\
\hline
 Participant 10 & \cellcolor{gray}0.0\% & \cellcolor{green}62.5\% & \cellcolor{red}37.5\% \\
 \hline
 \textbf{Mean}&\cellcolor{lightgray} 26.4\%&\cellcolor{lightgray}46.0\%&\cellcolor{lightgray}27.1\%\\
 \hline
\end{tabular}
\caption{Swing Speed Proportions for the Visualisation Test, showing the percentage of swings at varying speeds ($>$40 mph, 25 mph$<$$>$40 mph and $<$25 mph). Red indicates a worse performance and green indicates a better performance under this condition. Grey indicates an unchanged performance.}
\label{fig:Table8}
\end{table}

\begin{table}[H]
\centering
\begin{tabular}{ |>{\centering\arraybackslash}p{3.5cm}|>{\centering\arraybackslash}p{3.5cm}|>{\centering\arraybackslash}p{3.5cm}|>{\centering\arraybackslash}p{3.5cm}| }
 \hline
 \multicolumn{4}{|c|}{\cellcolor{lightgray}\large Statistical Analysis for Accurate Shots at varying swing speeds} \\
 \hline
 \bf{Participant}& \bf{Mean Difference (V - B)} &\bf{t-statistic}&\bf{p-value}\\
 \hline
 Greater than 40 mph & 2.9\% & 0.5791 & 0.5767\\
 \hline
 Between 25 mph and 40 mph & 7.1\% & 1.1631 & 0.2747 \\
 \hline
 Less than 25 mph & -10.4\% & -1.6073 & N/A \\
 \hline
\end{tabular}
\caption{Statistical analysis for accurate shots played at varying swing speeds.}
\label{fig:Table9}
\end{table}

Table 10 displays a comparison of the number of shots played per participant at a swing power greater than 75\% under both conditions. There was a mixed result shared between the participants as five had played more shots above 75\% in the visualisation test, four with less shots at this power level, and one with an unchanged number of shots. This is an interesting result in the context of visual data influence as some players attempted more powerful shots whereas others attempted to play more controlled shots after seeing metrics.

\begin{table}[H]
\centering
\begin{tabular}{ |>{\centering\arraybackslash}p{3.5cm}|>{\centering\arraybackslash}p{3.5cm}|>{\centering\arraybackslash}p{3.5cm}|>{\centering\arraybackslash}p{3.5cm}| }
 \hline
 \multicolumn{4}{|c|}{\cellcolor{lightgray}\large Number of shots at Swing Power $>$ 75\%} \\
 \hline
 \bf{Participant} & \bf{Baseline} & \bf{Visualisation} & \bf{Change} \\
 \hline
 Participant 1 & 0 & 6 & \cellcolor{green}+6 \\
\hline
 Participant 2 & 14 & 0 & \cellcolor{red}-14 \\
\hline
 Participant 3 & 8 & 16 & \cellcolor{green}+8 \\
\hline
 Participant 4 & 12 & 2 & \cellcolor{red}-10 \\
\hline
 Participant 5 & 1 & 4 & \cellcolor{green}+3 \\
\hline
 Participant 6 & 15 & 11 & \cellcolor{red}-4 \\
\hline
 Participant 7 & 13 & 21 & \cellcolor{green}+8 \\
\hline
 Participant 8 & 0 & 0 & \cellcolor{gray}0 \\
\hline
 Participant 9 & 4 & 1 & \cellcolor{red}-3 \\
\hline
 Participant 10 & 0 & 6 & \cellcolor{green}+6\\
 \hline
\end{tabular}
\caption{The number of shots played at a Swing Power greater than 75\%. Five participants played more shots above 75\% in the visualisation test, four played less shots at this power level, and one had an unchanged number of shots. Red indicates the participants with a lower number of shots in the visualisation test and green indicates the participants with a higher number of shots.}
\end{table}

\subsection{Qualitative Results}
\subsubsection{Key Findings}
\begin{itemize}
    \item 9/10 participants enjoyed the experience.
    \item Minimal impact from the sleeve
    \item 6/10 participants reported minimal discomfort from the VR equipment.
    \item The visual window distracted participants greatly.
    \item Some participants used the visual data as a guide to control their swings.
    \item Other participants used the data as a target to generate swings of larger values, without considering accuracy.
    \item A majority felt they improved their skills and understanding of tennis.
\end{itemize}

\section{Discussion}

\subsection{Performance}
The results in Tables 1 and 2 summarises the change in the percentage of accurate shots and the change in the number points scored. The degree of performance improvement shown in Table 1 is a varying for four participants. Further analysis of the four players revealed a significant reduction in the number of shots missed and of those that were out, showing a significant improvement of player performance in all aspects of the game with reduced errors.
Although the results do not show an improvement in shot accuracy for the remaining players, it remains that intrinsic improvement can be seen in certain aspects of each player, such as a reduction in missed shots and an increase number of accurate shots. As six participants increased the number of shots that won a point in the visualisation condition, there is consideration of the impact of the visualised data on these players. The results speculate that the data shown gave players a direct understanding of how their shots correlate with winning points. 
\vspace{3mm}

Analysing these two tables, contrasting results are seen. Some players highlighted red in Table 1 are highlighted green in Table 2. For example, Participant 1 had a lower percentage of accurate shots in the visualisation test, but simultaneously managed to score more points. This is also the case for Participants 7 and 9. In contrast, one participant managed to score less points while improving accuracy: Participant 10. These conflicting results indicate that the visualised data offered some participants more comprehensive data on which swings provided higher quality shots, rather than simply playing an accurate shot. Responses from the questionnaire highlight the reasoning for players.

Nevertheless, an important limitation of this study is the within-subjects design, where all participants completed the baseline test before the visualisation test. This sequence introduces a potential confounding learning effect, as improvements seen during the VR visualisation phase might be partially attributable to practice gained during the baseline phase, rather than solely due to the visual feedback. Future studies could mitigate this by including a control group that undergoes two sessions without visualisation.

\subsection{Speed Range and Performance Correlation}
The relationship between swing speed and performance was particularly informative. Most participants (7/10) increased the percentage of shots with speeds exceeding 40 mph during the visualisation test, demonstrating that seeing their metrics prompted attempts at more powerful shots. However, this shift toward higher speeds did not universally improve accuracy. Instead, the greatest accuracy improvements were observed in the moderate speed range (25-40 mph), suggesting that visualization helped players discover their optimal performance zone.
This finding aligns with participant feedback that the visual metrics helped them find a "sweet spot" between power and accuracy. One participant noted they were "more aware of swing power vs accuracy and where the sweet spot was," reinforcing Wu et al.'s \cite{Wu2021SPinPong} assertion that visualised performance data can lead to targeted technique adjustments.

\subsection{Cognitive Ergonomics and User Experience}
The questionnaire results revealed important insights into player psychology and cognitive engagement. While 90\% of participants enjoyed the experience, the visualisation interface presented challenges. The "floating" nature of the visual window was described as unnatural and distracting, with one participant noting it "didn't look like it was part of the game." This highlights the importance of seamless integration of data visualisation within the virtual environment.
The learning curve associated with VR was also evident, with several participants indicating they would "need to try the system a few times" to develop competency. This suggests that performance improvements might be more pronounced with extended exposure to the system. Importantly, the wearable device itself created minimal distraction, with participants unanimously reporting little attention paid to the sleeve during gameplay.

\section{Conclusion}
This study has outlined the capabilities of the impact of data visualisation on novice tennis players. The sensor device prototype successfully collected swing data and the system developed to help novice tennis players visualise the data was successful in displaying swing speed and swing power values in real-time. The study has shown that real-time visual feedback can not only provide insightful information to users, but can also influence them to make more conscious decisions in real-time to adjust technique, playing style and other contributing factors in order to achieve a better performance. The influence of the visualised data is clearly shown in both quantitative and qualitative results. 
\vspace{3mm}

Although a slight majority of the participants (6/10) were unable to statistically produce a higher total of accurate shots under the visualisation condition, most of the participants (7/10) had adjusted their swings to produce more accurate shots at an optimum speed range between 25 mph and 40 mph. In addition, 6/10 participants played a higher number of shots that won a point under the visualisation condition. For some players, the visualised data enabled them to consciously analyse their swings, and reproduce those that were accurate. This demonstrates the positive influence of the visualised data which resulted in more cognitive awareness. The visualised data provided users with direct feedback, establishing a basis upon which they could learn how to control their swings to achieve more accurate shots.
\vspace{3mm}

The positive outcomes observed, particularly in helping novice players identify optimal performance parameters, suggest that such technology could bridge the gap between casual gaming and serious sports training. This has important implications for how sports skills might be developed in the future, potentially democratizing access to advanced training techniques that were previously available only to elite athletes with professional coaching.
As VR technology becomes more accessible and widespread, the methodology and findings from this study provide valuable insights for developers, coaches, and sports scientists looking to create effective training tools. The integration of quantitative performance metrics with immersive experiences represents a promising approach to enhancing motor learning and skill development across various sporting domains.
\vspace{3mm}

The use of virtual reality to aid tennis training has provided a resounding majority of the participants with a proactive and enjoyable experience. These findings represent a promising development in the application of immersive technology and agrees with the body of evidence explored in sports training literature, encouraging future research in this area and other related fields.

\medskip

\medskip

%
\bibliographystyle{MSP}
\bibliography{references}

\end{document}